\documentclass[conference]{IEEEtran}
\IEEEoverridecommandlockouts
\usepackage{cite}
\usepackage{amsmath,amssymb,amsfonts}
\usepackage{algorithmic}
\usepackage{graphicx}
\usepackage{textcomp}
\usepackage{xcolor}
\def\BibTeX{{\rm B\kern-.05em{\sc i\kern-.025em b}\kern-.08em
    T\kern-.1667em\lower.7ex\hbox{E}\kern-.125emX}}
    
\usepackage{multirow}
\usepackage{booktabs}    
\usepackage{float}   
\usepackage{caption}
\usepackage{subcaption}
\PassOptionsToPackage{hyphens}{url}
\usepackage{url}
\usepackage{hyperref}
\begin{document}
    
\title{Automated Recovery of Issue-Commit Links
Leveraging Both Textual and Non-textual Data}

\author{\IEEEauthorblockN{Pooya Rostami Mazrae}
\IEEEauthorblockA{
\textit{Computer Engineering Department} \\
\textit{Sharif University of Technology}\\
prostami@ce.sharif.edu}
\and
\IEEEauthorblockN{Maliheh Izadi}
\IEEEauthorblockA{
\textit{Computer Engineering Department} \\
\textit{Sharif University of Technology}\\
maliheh.izadi@sharif.edu}
\and
\IEEEauthorblockN{Abbas Heydarnoori}
\IEEEauthorblockA{
\textit{Computer Engineering Department} \\
\textit{Sharif University of Technology}\\
heydarnoori@sharif.edu}
}
\maketitle

\begin{abstract}
An issue report documents the discussions around required changes in issue-tracking systems, while a commit contains the change itself in the version control systems.
Recovering links between issues and commits can facilitate many software evolution tasks such as bug localization, defect prediction, software quality measurement, and software documentation.
A previous study on over half a million issues from GitHub
reports only about 42.2\% of issues are manually linked by developers to their pertinent commits.
Automating the linking of commit-issue pairs can contribute to the improvement of the said tasks.
By far, current state-of-the-art approaches for automated commit-issue linking suffer from low precision, leading to unreliable results, sometimes to the point that imposes human supervision on the predicted links.
The low performance gets even more severe when there is a lack of textual information in either commits or issues.
Current approaches are also proven computationally expensive.

We propose \textit{Hybrid-Linker}, an enhanced approach that overcomes such limitations by exploiting two information channels; (1) a non-textual-based component that operates on non-textual, automatically recorded information of the commit-issue pairs to predict a link, and (2) a textual-based one which does the same using textual information of the commit-issue pairs.
Then, combining the results from the two classifiers, Hybrid-Linker makes the final prediction.
Thus, every time one component falls short in predicting a link, the other component fills the gap and improves the results.
We evaluate Hybrid-Linker against competing approaches, namely FRLink and DeepLink on a dataset of 12 projects.
Hybrid-Linker achieves 90.1\%, 87.8\%, and 88.9\% 
based on recall, precision, and F-measure, respectively. 
It also outperforms FRLink and DeepLink
by 31.3\%, and 41.3\%, regarding the F-measure.
Moreover, the proposed approach 
exhibits extensive improvements in terms of performance as well. 
Finally, our source code is publicly available.
\end{abstract}
\begin{IEEEkeywords}
Link Recovery, Issue Report, Commit, Software Maintenance, Machine Learning, Ensemble Methods
\end{IEEEkeywords}

\section{Introduction} \label{introduction}

\par 
Issues and commits are two software artifacts commonly used for various tasks
in software hosting platforms such as GitHub, Jira, and Bugzilla.
Issue reports encapsulate user discussions around different aspects of a software, 
as a sort of documentation.
Commits contain source code changes 
required to fix bugs, add features, improvements, etc discussed in the issues. 
Issues are usually reported in bug-tracking systems such as Bugzilla or Jira, 
on the other hand, 
corresponding commits are stored in version control systems such as GitHub~\cite{sun2017improving}.
There are also cases that they are both maintained in one system.
When a developer commits a change in a project,
it is a good practice to mention the issue in the commit 
to document the relationship between the two.
However, it is seldom the case 
due to the deadline's pressure, lack of motivation, etc.~\cite{sun2017improving}. 
To quantify the prevalence of missing issue-commit links, 
Ruan et al.~\cite{ruan2019deeplink} analyzed over half a million issues from GitHub. 
They report only 42.2\% of issues were linked to corresponding commits.
Recovering issue-commit links is deemed important 
for improving bug prediction solutions~\cite{le2015rclinker,ruan2019deeplink}, 
bug assignment~\cite{anvik2006should}, 
feature location techniques~\cite{dit2013feature},
and other software maintenance tasks. 
It can also be used to evaluate software maintenance efforts and quality~\cite{sun2017frlink}. 
Thus, an automated method for recovering links between
issues and their corresponding commits can be of high value.

\par The first challenge for such an approach
is to use a proper dataset of True and False Links between issues and commits. 
True Links are the correct links between issues and their related commits.
All the other combinations of links can be considered False Links.
Current approaches build these links manually.
This affects the reliability of results.
Moreover, some issues have more than one related commit.
An automatic solution to recovering True Links 
should be able to handle these relationships.
Another important aspect is the performance of proposed approaches.
Current studies mostly focus on the precision and recall scores of the predictions.
However, the prediction time and complexity of the models are also important.

\par In this work, we introduce a novel approach,
named \emph{Hybrid-Linker} 
to address the above-mentioned problems. 
Hybrid-Linker exploits both textual and non-textual data to achieve higher performance. 
Textual information includes the issue title, description, code difference, and commit messages.
Non-textual information consists of various characteristics of an issue and commits,
such as the author of an issue, 
the committer, commit time, 
type of an issue (bug, feature, task),
and state of a issue (open, closed, or resolved).  
We first identify all the relevant information 
and then perform feature engineering to extract the most important ones.
The reason for incorporating non-textual data is 
to enable Hybrid-Linker to exploit this knowledge 
when there is little textual information available (e.g., there is no commit messages), 
or there are few similarities between 
the description of an issue and textual information of a commit. We train a hybrid model consisting of two classifiers and 
a module to achieve the best linear composition of these classifiers. 
The non-textual component is an ensemble of two classifiers.
The textual component is created 
using TF-IDF word embeddings and a single classifier.
We evaluated Hybrid-Linker against two baseline methods, 
FRLink and DeepLink for $12$ projects with different characteristics. 
In summary, our contributions are as follows:
\begin{itemize}
    \item Proposing an automatic approach, called \textit{Hybrid-Linker}, 
    for recovering the links between issues and commits 
    using a hybrid model of classical classifiers.
    \item Our results show that Hybrid-Linker outperforms the competing approaches, FRLink and DeepLink,
    by 31.3\%, and 41.3\% respectively, regarding the F-measure.
    Moreover, our proposed approach shows extensive improvements in terms of required training time.  
    \item Finally, we release our source code and data publicly.\footnote{\url{https://github.com/MalihehIzadi/hybrid-linker}}
\end{itemize}
    
\section{Motivating Example} \label{motivation example}
\par
\begin{figure}[h]
     \centering
     \begin{subfigure}[b]{\linewidth}
         \centering
         \includegraphics[width=\textwidth]{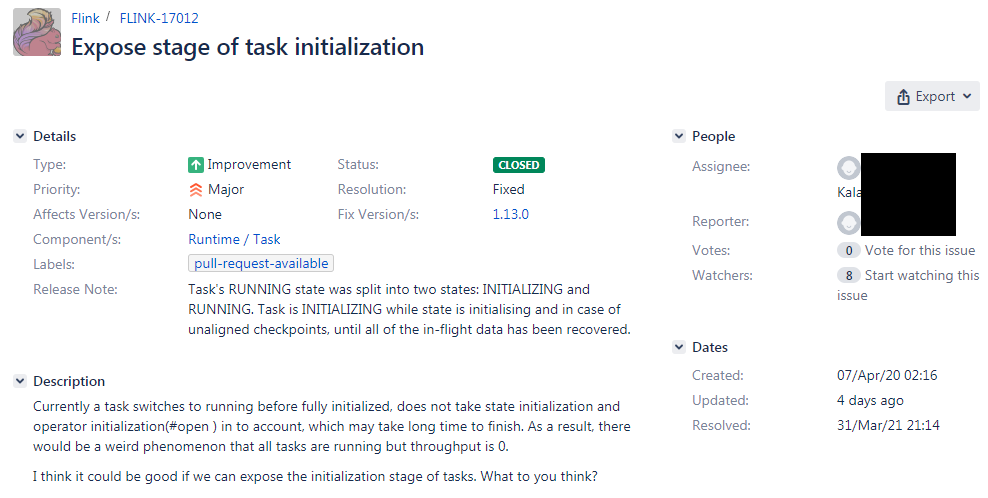}
         \caption{Example of an issue}
         \label{figure:issue example}
     \end{subfigure}
     
     \begin{subfigure}[b]{\linewidth}
         \centering
         \includegraphics[width=\textwidth]{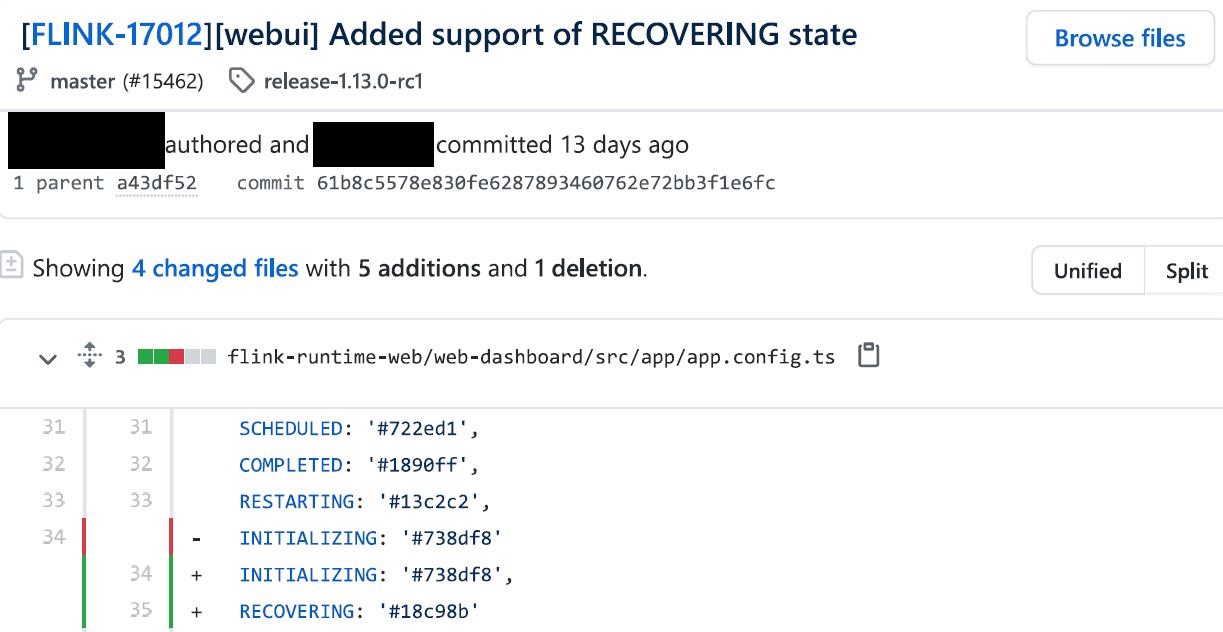}
         \caption{Example of part of a commit}
         \label{figure:commit example}
     \end{subfigure}
        \caption{Example of a True Link between an issue and a commit}
        \label{figure: example of issue and commit}
\end{figure}

 Here, we illustrate an example as the motivation 
for enhancing automatic link recovering approaches between issues and commits.
Figure~\ref{figure:issue example} is an example of an issue\footnote{\url{https://issues.apache.org/jira/browse/FLINK-17012}}. 
Figure~\ref{figure:commit example} shows an example of a commit\footnote{\url{https://bit.ly/2PCsQu6}} 
related to the above-mentioned issue. 
The issue and the commit are selected from \emph{Flink} project. 
Apache Flink is an open-source, unified stream-processing and batch-processing framework 
developed by the Apache Software Foundation. 
An issue has different fields like type, status, release note, description, created date, updated date, and resolved data.
A commit contains commit message, committer ID, author ID, name of changed files, and Diff of changed files. 
Note that other information such as comments and code snippets attached to some issues do not always exist.
\par As shown, there is no compelling similarity 
between the text of issue description, its release note 
and the respective commit message. 
Due to lack of similarity in textual information of this issue and commit,
FRLink approach fails to discover the True Link between them~\cite{sun2017frlink},
Moreover, DeepLink~\cite{ruan2019deeplink} approach 
also struggles to identify this link
as there is no code snippet in the description section of the issue.
Thus, DeepLink will find little semantic relation 
between the issue and the source code in this commit.
To address these problems, 
we propose to extract knowledge 
from both textual and non-textual channels of issues and commits.
Then combine this information in a hybrid model to train stronger link recovery models.

\section{Proposed Approach} \label{proposed approach}

\begin{figure}
    \centering
    \includegraphics[width=0.95\linewidth]{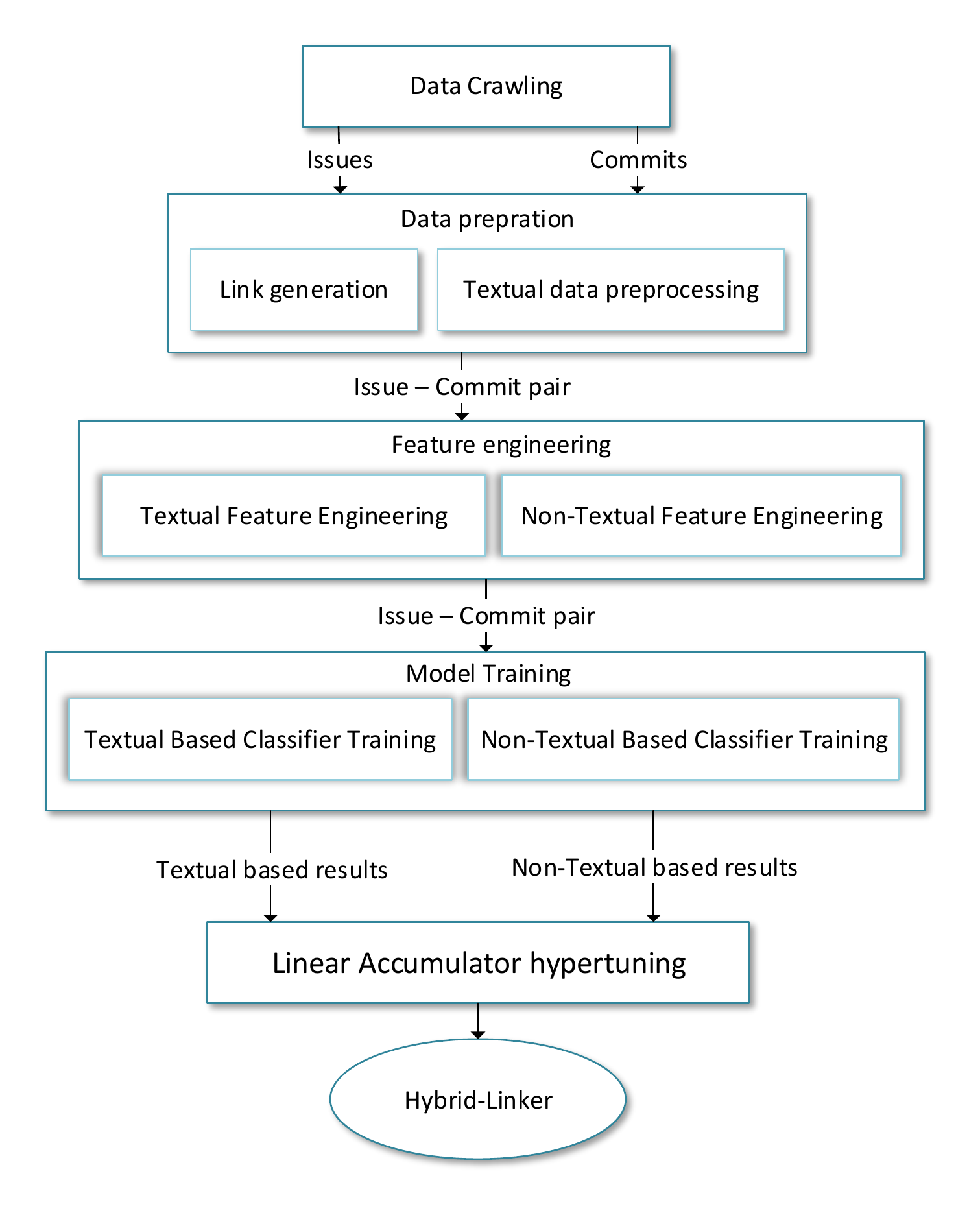}
    \caption{Overview of the proposed approach}
    \label{figure:approach_overview}
    \vspace{-4mm}
\end{figure}

In this section, we present the main steps of our approach, namely: 
(1) \emph{data crawling},
(2) \emph{data preparation},
(3) \emph{feature engineering},
(4) \emph{model training}, and
(5) \emph{linear accumulator hyper-tuning}.
\autoref{figure:approach_overview} illustrates an overview of the approach 
and the following provides a detailed description of each of the five aforementioned steps.

\subsection{Data Crawling} \label{data crawling}

While we utilize a dataset from Claes et al. \cite{claes202020}, this dataset does not satisfy our needs for the approach.
More specifically, we aim to incorporate the \texttt{code diff} data of the commits into the solution, which is missing from the shared dataset due to the large volume of such data.
To be able to crawl the excessively large code diff data, 
we first reduce the number of projects to a sample of $12$ projects.
Then, for the sampled projects, we crawl the missing data from the projects' code bases to complement the dataset.
Moreover, the dataset is provided in a segmented state (separated commits and issues). To uniformize it, for each project, we concatenated segments of data.

\subsection{Data Preparation} \label{data preparation}

In this step, we prepare the dataset from two aspects.
In the \emph{link generation} process, 
we generate the data points, i.e., the issue-commit pairs.
These data points are assigned with \emph{True Link} labels
when the link between the issue and its paired commit is in place and \emph{False Link} otherwise, i.e., the issue and commit are irrelevant.
We also perform \emph{textual data preprocessing} techniques 
on the textual data of issues and commits separately to prepare them for feature engineering.
The following elaborates on these two data preparation steps.

\subsubsection{Link Generation}\label{link generation}

In the dataset, there are instances of issues and commits 
that are already linked by the developers, 
which means they are established and validated as True Links.
We take such pairs of issue-commits as data points with the True Link label.
To train the classifiers, 
we need to provide the model with data points labeled as False Link as well.
However, such data points are not explicitly included in the dataset.
Thus, we need to generate them such that their label, False Link, is ensured.
To do so, we pair the commits that are already linked to an issue 
by the developers with any issue other than the ones they are already linked to.
Since the commits only make a True Link with the issues selected by the developers, 
pairing them with any other issue makes a False Link.

However, taking all the generated links as False Link data points 
makes the dataset extremely imbalanced.
To perceive the severity of this problem in the context of a project, 
consider a project with $c$ number of commits that are already linked to an issue by the developers.
If the same project contains $i$ number of issues, 
$c$ of which are already linked to the aforementioned commits.
For each commit already linked to an issue, 
there are $i - 1$ issues each posing as a potential False Link pair for the commit.
Hence, the number of False Link data points adds up to $c*(i-1)$ issue-commit pairs.

To address this issue, 
we use the criteria used by previous work~\cite{sun2017frlink,ruan2019deeplink} 
to generate the False Links.
Thus we compare the two relevant submission dates of a commit with the three date attributes in an issue report
and construct a new False link 
if the commit is submitted seven days before or after any of these three issue attributes. 
Table \ref{table:projects information summary} 
presents the information of the resultant dataset.
As shown in this table, even after employing the above-mentioned criteria,
the number of False Links is much larger than those of True Links for all projects.
To further alleviate the imbalanced nature of this dataset, 
we apply a common data balancing technique.
More specifically, we randomly select the same number of False Links as the True Links in each project,
to provide our classifier with completely balanced datasets.
\begin{table}
    \caption{Selected projects' information}
    \begin{center}
     \begin{tabular}{c |c c |c | p{8mm} p{8mm}} 
     \toprule
     \textbf{Project} & \textbf{\#Issues} & \textbf{\#Commits} & \textbf{\#Stars} & \textbf{\#True Links} & \textbf{\#False Links} \\ [0.15ex] 
     \midrule
     Beam & 9133 & 28824 & 4300 & 5750 & 1505559\\ 
     Flink & 15655 & 26517 & 14500 & 14472 & 4850083\\
     Freemarker & 127 & 4235 & 604 & 177 & 382\\
     Airflow & 6511 & 10196 & 18800 & 5295 & 1030733\\
     Arrow & 7509 & 6179 & 6400 & 5252 & 1006904\\
     Netbeans & 3705 & 19184 & 1500 & 1369 & 129639\\
     Ignite & 12495 & 32930 & 3500 & 9997 & 2087327\\
     Isis & 2264 & 15284 & 580 & 8486 & 260259\\ 
     Groovy & 9117 & 30478 & 3900 & 8851 & 457876 \\ 
     Cassandra & 15413 & 31491 & 6300 & 146 & 40415\\ 
     Ambari & 25162 & 38872 & 1400 & 35589 & 8891881\\ 
     Calcite & 3740 & 6934 & 2100 & 3058 & 201106\\ [0.15ex] 
     \bottomrule
    \end{tabular}
    \end{center}
    \label{table:projects information summary}
\end{table}

\subsubsection{Textual Data Preprocessing}\label{textual data preprocessing}

The resultant dataset contains textual and non-textual data on issues and commits from the sampled projects.
The textual data contains both natural language text such as \emph{issue title}, \emph{issue description} and \emph{commit message}, and the \emph{code diff}.

We first clean and preprocess the input textual data.
We perform the three commonly-used strategies of tokenizing, 
removing stop words, 
and stemming on the natural language text data as the preprocessing step.
These preprocessing actions not only reduce the vocabulary size,
which in turn makes the feature set a compact one, 
but also they integrate different forms of words 
by replacing them with their roots.

As for the diff data, while they do include multiple lines of code per sample, 
only the identifiers, i.e., method and variable names, 
carry valuable information about the changes in a commit.
That is because many of the keywords and commonly used method calls in the diff appear all over the code without indicating the purpose of the code snippet, 
while identifiers, 
if named according to the software development guidelines, 
refer to their purpose, role, and/or task.
Hence, we aim to extract only the identifiers through the use of code term patterns.
The code term patterns we employ are the ones 
previously used by Sun et al.~\cite{sun2017frlink} and Ruan et al.~\cite{ruan2019deeplink} 
(defined in \autoref{table:code patterns}).

\begin{table}
    \caption{Code term patterns introduced in FRLink~\cite{sun2017frlink}{}}
    \centering
     \begin{tabular}{c c c} 
     \toprule
     \textbf{Type} & \textbf{Example} & \textbf{Regular Expression} \\ [0.15ex] 
     \midrule
     C\textunderscore notation & OPT\textunderscore INFO & [A-Za-z]+[0-9]*\textunderscore.* \\ 
     Qualified name & op.addOption & [A-Za-z]+[0-9]*[\textbackslash.].+ \\
     CamelCase & addToList & [A-Za-z]+.*[A-Z]+.* \\
     UpperCase & XOR & [A-Z0-9]+ \\
     System variable & \textunderscore cmd & \textunderscore+[A-Za-z0-9]+.+ \\
     Refrence expression & std::env & [a-zA-Z]+[:]\{2,\}.+ \\ [0.15ex] 
     \bottomrule
    \end{tabular}
    \label{table:code patterns}
    \vspace{-4mm}
\end{table}

\subsection{Feature Engineering} \label{feature engineering}

We leverage both textual and non-textual data 
to improve the results of the True Link prediction task.
However, the features in the textual and non-textual feature vectors are not equally valuable in terms of being determinative of a True Link.
In the textual data context, there might be distinct words throughout the dataset 
that appear in a significant number of the data points.
This signifies that they are simply common tokens throughout the project 
and can not be considered as the indicator of the subject of a commit.
In the non-textual context, this problem manifests itself 
in highly correlated columns of data or even almost identical ones.
There is also the case of almost empty columns in which the data is null-valued more often than not.

This makes the feature vector unnecessarily extensive, 
which makes it harder for the classifier models to converge 
due to the multitude of parameters they are to optimize.
Even if the classifier does converge and yield better results with such data included, 
the improvement is negligible and unjustifiable when evaluated against the computational costs.
For these reasons, we perform a feature engineering process on both the textual and non-textual data 
to reduce the size of feature vectors 
and keep both the performance of the solution and the computational costs of the model optimal.
The feature engineering processes performed are detailed in the following.

\subsubsection{Textual Feature Engineering} \label{textual feature engineering}

We employ the widely used data modeling technique, \emph{TF-IDF}, 
which captures the importance of the tokens based on probabilistic measures over the dataset.
This data modeling technique computes the term frequency of each term (token) in each document and document frequency of each term over the dataset and combines the former with the inverse of the latter to calculate a measure of importance for each term in the dataset. 
The higher the value of the TF-IDF measure for a term, the more probable the term is to contribute to the label prediction.

We apply the TF-IDF technique on natural language textual data of commits and issues and the code diff textual data separately. 
This generates three vectors of TF-IDF features for each data point.
Then, we concatenate the resultant vectors and construct one textual feature vector per data point.
The reason for such an approach is that the information and vocabulary in issues and commits inherently differ.
Moreover, according to our experiments, this approach leads to higher accuracy compared to when we combine the three input data first and then apply the TF-IDF technique on the concatenated text.
Our textual features are a commit's \textit{Message} and \textit{Diff}, and an issue's Summary and Description. 
Note that we have also applied Word2Vec and Doc2Vec techniques, 
however, TF-IDF embeddings produced the best result.
\subsubsection{Non-textual Feature Engineering} \label{non-textual feture engineering} 

In the case of highly correlated columns, 
reducing them to a single column can improve the computational costs of the model training process 
by limiting the number of optimization parameters, 
while preserving the performance of the classifier.
We extensively inspect the dataset for such strongly correlated columns 
among commit features an issue features by calculating the similarity and correlation among the columns of similar types.
We discover that among the issue data columns, 
over $99\%$ of the data points have the same value as the \emph{reporter} and the \emph{creator}.
This makes these two columns practically duplicate.
Hence, we drop one and keep the other.
We also detect that over $65\%$ of the data points 
have the same \emph{author} and \emph{committer} in their commit data columns.
As we believe a similarity of $65\%$ is not high enough to justify the omission of one of the columns, 
so we keep both columns in the dataset.

Since the categorical data will be converted to a one-hot model, 
each distinct value in the categorical data column will serve as a Boolean feature.
Thus, the multitude of distinct values in a categorical column 
results in an over-complicated feature vector with too many features 
but very few true points, also known as a sparse matrix.
To avoid such an occurrence, we study the histograms of the categorical data 
and discovered that due to differences in labeling style across projects, 
the distinct values of the \emph{commit\textunderscore status} and \emph{issue\textunderscore type} columns 
can be mapped to two reduced sets of values.
For commits, status values was reduced from a set of $11$ statuses 
to three main categories of \emph{open}, \emph{closed}, and \emph{resolved}. 
We also reduced the set of $15$ distinct values of issue types 
to three main categories of \emph{task}, \emph{new feature}, and \emph{bug}. 

While there are two columns of highly correlated dates for issues, 
namely the \emph{create\textunderscore date} and the \emph{update\textunderscore date}, 
these dates prove as important features for the prediction of True Links.
The same goes for the \emph{author\textunderscore time\textunderscore date} and the \emph{commit\textunderscore time\textunderscore date} among the data of the commits.
We keep these columns intact to the dataset.

Finally, we drop the columns which have a significant number of null values.
After one-hot transformation of the categorical data, 
we calculate the correlations among all the columns, 
including the label column, for issues and commits separately. 
This is to verify that there are no correlations among the features and target column.
After it is verified that the dataset is not biased, 
the resultant commit and issue feature vectors are concatenated 
to compose a single feature vector for each data point.
Our non-textual features are 
\textit{commit time}, \textit{authoring time}, \textit{author hash}, and \textit{commit hash} of commits. 
We also include \textit{updated date}, \textit{created date}, \textit{status} (closed, open, resolved), \textit{issue type} (bug, new feature, task) and \textit{creator hash} from issue reports.

\subsection{Model Training} \label{model training}

We aim to keep the classifier model simple 
to lower the computational costs of training and prediction.
We believe one can improve the prediction accuracy of these models
by augmenting the input data.
To do so, we leverage both textual and non-textual data on the commits and issues and construct a hybrid model by training two classifiers, one that operates on textual data
and calculates the probability of labels, 
and another one that does the same using non-textual data.

\subsubsection{Textual Classifier Model} \label{textual classifier model}

As the textual classifier component, we train multiple classification models, namely a Decision Tree (DT), a Gradient Boosting (GB), a Logistic Regression (LR), and a Stochastic Gradient Descent (SGD) model to choose the model with the best performance among them.
We feed these models the resultant feature vectors from Section \ref{textual feature engineering} and train them.
The trained models take as input the processed vector of textual data and predict a label, either True or False Link for an issue-commit pair.

\subsubsection{Non-textual Classifier Model} \label{non-textual classifier model}

\begin{figure}
    \centering
    \includegraphics[width=0.99\linewidth]{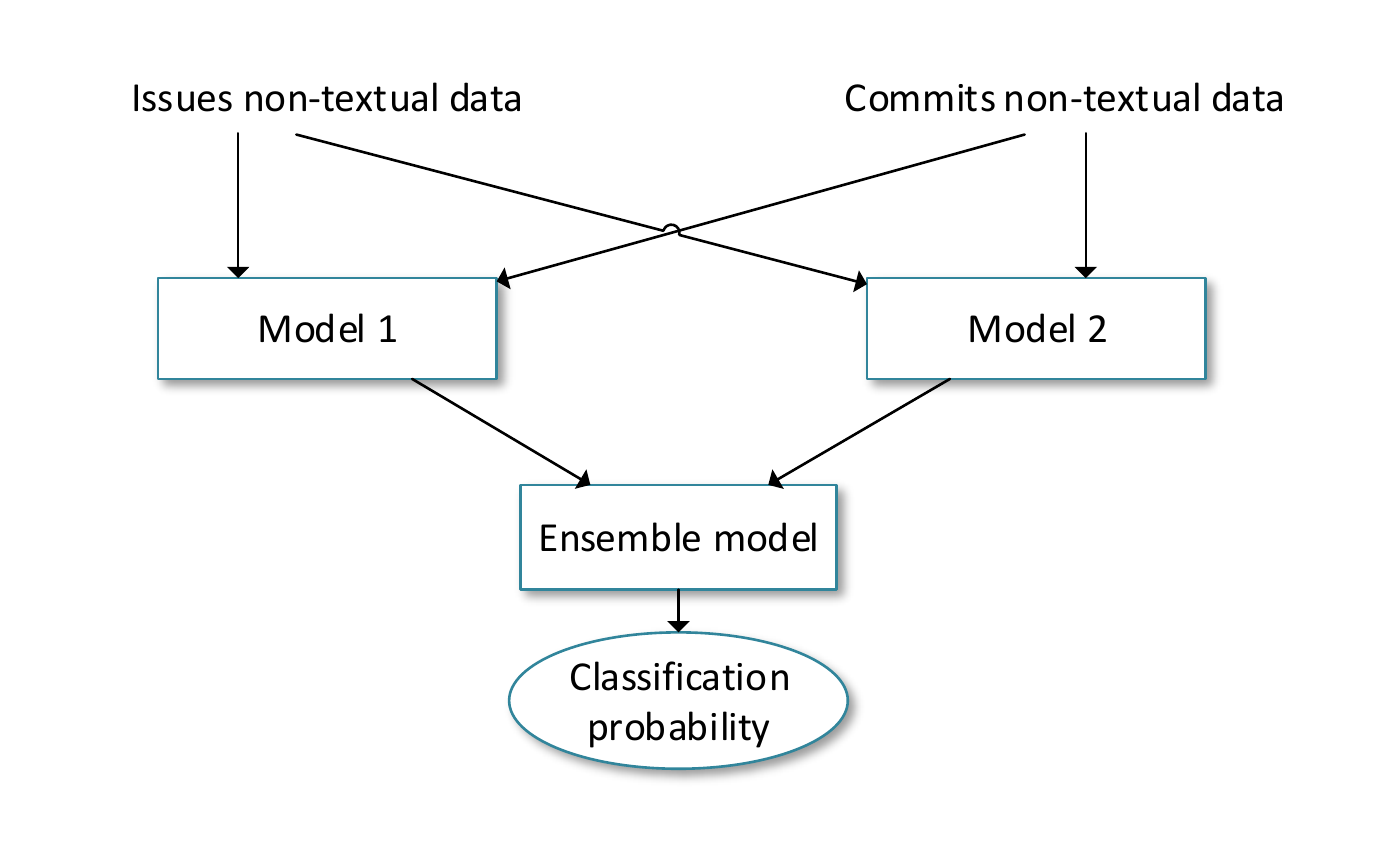}
    \caption{Overview of ensemble models for non-textual classifier component}
    \label{figure:non-textual_approach_overview}
    \vspace{-4mm}
\end{figure}

Here, we also use single and ensemble models to achieve the best results~\cite{pavlyshenko2018using}.
As simple classifier models, we train a Gradient Boosting~\cite{friedman2002stochastic}, a Naive Bayes (NB)~\cite{ting2011naive}, a Generalized Linear (GL), a Random Forest (RF)~\cite{breiman2001random}, and a XGBoost~\cite{chen2016xgboost} model.
To construct the ensemble models, following the overview illustrated in \autoref{figure:non-textual_approach_overview}, 
we combine the models and make four ensemble models accordingly.
The ensemble models are a $RF+GB$, a $GB+XGBoost$, a $RF+XGBoost$, and a $RF+GB+XGBoost$ combinations.

\subsection{Linear Accumulator Hyper-tuning} \label{linear accumulator hyper-tuning}

The last step in our approach is to combine the predictions of the two classifier components and generate the final prediction on the label of the data points.
To do so, we combine the predicted probability of a data point 
being a True Link from the two components 
and with a linear accumulator function, defined as the following:
\begin{equation}\label{eq:alpha}
    P_f = \alpha \times P_{nt} + (1-\alpha) \times P_t
\end{equation}
in which $P_f$ is the final calculated probability of a commit-issue pair 
being a true-link, 
$P_{nt}$ is the probability of the same pair being a true-link 
according to the non-textual classifier component, 
and $P_t$ is the same probability according to the textual classifier component.

In \autoref{eq:alpha}, $\alpha$ is the hyper-parameter 
by which we tune the model to produce the best results 
tailored to the characteristics of each project.
To do so, we vary the value of $\alpha$ from $0.00$ to $1.00$ in $0.05$ steps.
The value $\alpha$ by which best results 
regarding the F1-score is yielded is taken as the optimal $\alpha$ value.

\section{Experimental Design} \label{experminet}

\subsection{Research Questions} \label{research questions}
\par We define three Research Questions (RQ) 
to measure the effectiveness of our proposed approach.
We review these questions in the following.
\begin{itemize}
    \item \textbf{RQ1}: \emph{Compared to the state-of-the-art approaches,
    how effective is our approach in recovering the missing links between issues and commits?}
    To answer this question, we evaluate our method's performance 
    using the \textit{20MAD} dataset (reviewed in the next section)~\cite{claes202020}. 
    We use $12$ projects from this dataset for training and testing our model.
    There are different approaches for commit-issue recovery. 
    We use two of the state-of-the-art models,
    namely FRLink~\cite{sun2017frlink} and DeepLink~\cite{ruan2019deeplink}, 
    to compare with the proposed approach.
    \item \textbf{RQ2}: \emph{How to combine the two components of the model to achieve the best outcome?}
    There are different ways to combine our two models (textual and non-textual). 
    With this question, we aim to identify the best method to build a hybrid model.
    \item \textbf{RQ3}: \emph{What is the effect of each component of the model on the outcome?}
    As our model is constructed of different components, 
    we assess the benefits of adding each through an ablation study.
    That is, we evaluate the model using each of the two components separately and then compare the results by those of the Hybrid-Linker.
\end{itemize}

\subsection{Data Selection} \label{data selection}

\par As the data of previous work were not publicly shared, we utilized the dataset
presented by Claes and Mantyla~\cite{claes202020} in the MSR conference, 2020. 
From the Apache projects, we chose $12$ based on two criteria; 
(1) having a repository with more than $500$ stars (to have good input data for training the models), and 
(2) having a diverse number of issues for different projects (to be fair). 
As of September 2020, the number of stars for the selected projects was in the range of $580$ to $18800$. 
The second criterion let us choose projects with different number of issues 
from small to large software projects. 
The number of issues among our projects 
range from a couple hundred to more than $25$K issues. 
To prepare this data for feeding our Machine-Learning-based models, 
we complement and transform the selected $12$ projects from the 20MAD dataset 
as explained in Sections~\ref{data crawling} and \ref{data preparation}.

\subsection{Evaluation Metrics} \label{evaluation metric}
\par As previously used in the related work,
we use three metrics of Precision, Recall, and F-measure to evaluate the performance of the approach~\cite{ruan2019deeplink,sun2017frlink}.
These metrics are calculated using the following equations.
\begin{equation}
    Precision=\frac{TP}{TP + FP}
\end{equation}
\begin{equation}
    Recall=\frac{TP}{TP + FN}
\end{equation}
\begin{equation}
    F1=\frac{2 \times Precision \times Recall}{Precision + Recall}
\end{equation}

\subsection{Experiment Setting} \label{experiment setting}
\par For preprocessing, we use Pandas~\cite{mckinney2011pandas} library.
For training the classifiers
in the non-textual component of our approach, 
we use H2O.ai~\cite{aiello2016machine} library. 
H2O benefits from distributed, in-memory processing which results in faster models. 
It is also able to manage hash data better than the Sci-Kit Learn~\cite{pedregosa2011scikit} library. 
This library can be used in different programming languages such as Python and R. 
We use the Python version and 
added a Java Runtime Environment for the backend. 
For the textual component, we use the Sci-Kit Learn library. 
It has great I/O which lets us use different types of data 
like Parquet and Pickle files simultaneously. 
\par We use five-fold cross-validation to evaluate the models more thoroughly. That is, we break the data into five parts randomly, 
choose one as the test set and use the others for training.
After repeating this process five times, iterating over the parts as the test set, we report the average as the result of evaluations.
This also helps with the generalizability of the approach and avoiding the overfitting problem.

\par To find the best parameters for the ensemble model in the non-textual component, 
we perform a Random Grid search for each project of the dataset. 
The result of the search indicate that 
(n\textunderscore trees=60, max\textunderscore depth=15, min\textunderscore rows=2, learn\textunderscore rate=0.1, learn\textunderscore rate\textunderscore annealing=1) 
for the Gradient Boosting model 
and (n\textunderscore trees=60, max\textunderscore depth=15, min\textunderscore rows=2, learn\textunderscore rate=0.1) 
for the XGBoost model are the best choices.

\par To find the best parameters for the Gradient Boosting in the textual component, 
we perform another Random Grid search for each project of the dataset. 
The result of the search indicate 
that (n\textunderscore estimators=300, max\textunderscore features=None, max\textunderscore depth=50 and learning\textunderscore rate=0.1) are the best choices.

\par To build the TF-IDF embedding vectors, 
we experiment with unigram, bigram, and union of unigram, bigram, and trigram word embedding. 
The best case is the union of unigram, bigram, and trigram 
as it finds all the important individuals and combinations of words. 
For TF-IDF embeddings, we set a maximum number of features to $10$K.

\par For all of our experiments, we used the same machine with $32$GB memory and a 4-core Intel i7-7700k 4.2G processor. 

\par The baselines here are FRLink~\cite{sun2017frlink} and DeepLink~\cite{ruan2019deeplink}
as they achieve the state-of-the-art results in the problem of
automatically recovering links between issues and their corresponding commits.
FRLink uses a set of features and complementary documents 
such as non-source documents to learn from relevant data for recovering links~\cite{sun2017frlink}.
They analyze and filter out irrelevant source code files 
to reduce data noise. 
On the other hand, 
\emph{DeepLink} uses a semantically-enhanced link recovery method
based on deep neural networks~\cite{ruan2019deeplink}. 
The authors apply a recurrent neural network 
on the textual information of issues and commits
for training their model.
They also disregard issue comments due to their length and noise. 
While DeepLink outperforms FRLink in terms of F-measure,
it achieves lower recall scores~\cite{ruan2019deeplink}.
Thus, we use both these techniques here as the baselines 
to compare our approach with.
We use the replication packages provided by Ruan et al.~\cite{ruan2019deeplink} 
for these two models\footnote{\url{https://github.com/ruanhang1993/DeepLink}}.
We slightly modified their input reader function to be able to read our data.
Moreover, we set all the parameters as specified in the original papers.

\section{Results} \label{results}

In this section, we answer the research questions by providing the results of the experiments.
We first, compare the performance of our proposed approach with the state-of-the-art ones.
Next, we review the results of our investigations on how to build a hybrid model.
Finally, we present the results of the ablation study to show the effectiveness and impact of each component of the proposed approach.

\paragraph{RQ1} \emph{Compared to the state-of-the-art approaches, how effective is our approach in recovering the missing links between issues and commits?}
To answer this RQ, we built our approach with two classifier components, a textual classifier and a non-textual one that each predict the probability of a issue-commit pair being a true-link.
We plugged multiple classifier models into each of the said components and chose the models with best performances as our proposed ones.

For the textual classifier component, we fed the concatenated TF-IDF vectors to four classifier models widely used for text classification purposes and study the results to determine the best performance among the models.
\autoref{table:textual_results} shows the outcome of the trained models.
The results indicate the best algorithm for classifying issue and commits linkage based on their textual data is the Gradient Boosting model.

For training a classifier on non-textual information, we experimented with well-known classical classifiers to identify the best classifier for our case. 
As seen in Table~\ref{table:non_textual_results} Gradient Boosting, Random Forest, and XGBoost have higher overall metrics results. 
Moreover, ensemble methods have been shown to outperform simple models.
Thus, we also opt for an ensemble model of the above algorithms to identify the best combination here. 
Based on results in \autoref{table:non_textual_results}, the ensemble of Gradient Boosting and XGBoost produce the best result for our non-textual data. 
Table~\ref{table:non_textual_results} reports 
the average score of precision, recall and F-measure for each model. 
\begin{table}
    \caption{The average performance of models on textual data.}
    \centering
    \begin{tabular}{l | c c c }
    \toprule
    \textbf{Project} & \textbf{Recall} & \textbf{Precision} & \textbf{F-measure} \\
    \midrule
    Decision Tree & 73.30\% & 73.58\% & 73.25\%\\
    Gradient Boosting & \textbf{78.27\%} & \textbf{78.38\%} & \textbf{78.25\%}\\
    Logistic Regression & 64.60\% & 65.08\% & 64.55\%\\
    Stochastic Gradient Descent & 63.67\%  & 63.91\% & 63.61\%\\ [0.15ex]   
    \bottomrule
    \end{tabular}
    \label{table:textual_results}
    \vspace{-3mm}
\end{table}

\begin{table}
    \caption{The average performance of models on non-textual data.}
    \begin{center}
     \begin{tabular}{c | c | c c c} 
    \toprule
    \textbf{Method} & \textbf{Algorithm} & Recall & Precision & F-measure \\ [0.15ex] 
    \midrule
     \multirow{5}{4em}{Simple Method} & Gradient Boosting(GB) & \textbf{100\%} & \textbf{96.34\%} & \textbf{85.50\%} \\ 
     & Naïve Bias & 100\% & 66.97\% & 73.77\% \\
     & Generalized Linear & 100\% & 88.98\% & 76.18\% \\
     & Random Forest(RF) & \textbf{100\%} & \textbf{98.25\%} & \textbf{84.55\%} \\
     & XGBoost & \textbf{100\%} & \textbf{98.67\%} & \textbf{86.14\%} \\ [0.15ex] 
    \midrule
     \multirow{4}{4em}{Ensemble Method} & RF + GB & 100\% & 98.29\% & 86.61\% \\ 
     & GB + XGBoost & \textbf{100\%} & 98.73\% & \textbf{87.81\%} \\
     & RF + XGBoost & 100\% & \textbf{99.27\%} & 87.10\% \\
     & RF + GB + XGBoost & 100\% & 98.81\% & 87.77\% \\ [0.15ex] 
    \bottomrule
    \end{tabular}
    \end{center}
    \label{table:non_textual_results}
\end{table}

The effectiveness of our proposed method is evaluated based on three metrics, namely Precision, Recall, and F-measure. 
Table~\ref{table:our approach vs Deeplink vs FRLink} presents a summary of the average performance of our approach across projects.
According to the results, our approach achieves an average of 
$90.14\%$ on Recall, $87.78\%$ on Precision, and $88.88\%$ of F-measure. 
Respectively, the lowest Recall is $84.41\%$ for \emph{Arrow} and the highest is $100\%$ for \emph{Cassandra} project. 
On the other hand, the lowest precision is $81.81\%$ for \emph{Cassandra} and the highest is $96.04\%$ for \emph{Ambari}.

\begin{table*}
    \caption{Performance of the models}
    \centering
     \begin{tabular}{c | c c c | c c c | c c c} 
    \toprule
    
    \multirow{1}{*}{} &
      \multicolumn{3}{c|}{\textbf{Hybrid-Linker}} &
      \multicolumn{3}{c|}{\textbf{DeepLink}} &
      \multicolumn{3}{c}{\textbf{FRLink}} \\
    \textbf{Project} & Recall & Precision & F-measure & Recall & Precision & F-measure & Recall & Precision & F-measure\\ [0.15ex] 
    \midrule
     Beam & 85.77\% & \textbf{86.22\%} & \textbf{85.99\%} & 82.63\% & 55.15\% & 66.15\% & \textbf{100\%} & 50.43\% & 67.05\%\\ 
     Flink & \textbf{91.91\%} & \textbf{89.69\%} & \textbf{90.79\%} & 43.98\% & 63.43\% & 51.94\% & 88.63\% & 61.80\% & 72.82\%\\
     Freemarker & 88.89\% & 91.42\% & 90.14\% & 95.83\% & \textbf{100\%} & \textbf{97.87\%} & \textbf{97.22\%} & 61.40\% & 75.26\%\\
     Airflow & 87.80\% & \textbf{85.72\%} & \textbf{86.75\%} & 44.54\% & 64.45\% & 52.67\% & \textbf{94.32\%} & 66.77\% & 78.19\%\\
     Arrow & 84.41\% & \textbf{83.71\%} & \textbf{84.06\%} & 16.85\% & 44.65\% & 24.47\% & \textbf{99.90\%} & 52.14\% & 68.52\%\\
     Netbeans & 88.84\% & \textbf{85.66\%} & \textbf{87.22\%} & 57.39\% & 73.56\% & 64.48\% & \textbf{92.93\%} & 62.65\% & 74.85\%\\
     Ignite & 90.82\% & \textbf{89.59\%} & \textbf{90.20\%} & 68.58\% & 70.16\% & 69.36\% & \textbf{100\%} & 50.71\% & 67.29\%\\
     Isis & 88.13\% & \textbf{89.84\%} & \textbf{88.98\%} & 47.78\% & 74.80\% & 58.31\% & \textbf{100\%} & 49.39\% & 66.12\%\\ 
     Groovy & 89.15\% & \textbf{87.79\%} & \textbf{88.47\%} & 47.65\% & 62.5\% & 54.07\% & \textbf{94.26\%} & 54.83\% & 69.33\%\\ 
     Cassandra & \textbf{100\%} & 81.81\% & \textbf{90\%} & \textbf{72.72\%} & \textbf{84.21\%} & 78.04\% & 100\% & 45.76\% & 62.79\%\\ 
     Ambari & 97.13\% & \textbf{96.04\%} & \textbf{96.58\%} & 87.50\% & 72.11\% & 79.06\% & \textbf{98.57\%} & 62.13\% & 76.22\%\\ 
     Calcite & 88.85\% & \textbf{85.89\%} & \textbf{87.34\%} & 55.58\% & 60.74\% & 58.04\% & \textbf{96.55\%} & 61.80\% & 75.36\%\\
    \midrule
     \textbf{Avg.} & \textbf{90.14\%} & \textbf{87.78\%} & \textbf{88.88\%} & \textbf{60.09\%} & \textbf{68.81\%} & \textbf{62.87\%} & \textbf{96.86\%} & \textbf{53.61\%} & \textbf{67.67\%}\\
     \textbf{Diff from Hybrid-Linker} & & & & \textbf{(-30.05\%)} & \textbf{(-18.97\%)} & \textbf{(-26.01\%)} & \textbf{(+6.72\%)} & \textbf{(-34.17\%)} & \textbf{(-21.21\%)}\\
     [0.15ex] 
    \bottomrule
    \end{tabular}
    \label{table:our approach vs Deeplink vs FRLink}
    \vspace{-4mm}
\end{table*}

We compare our approach with two of the competing models, namely FRLink and DeepLink. 
On average, our approach has $34.17\%$ higher precision and $21.21\%$ higher F-measure scores than FRLink. 
Although FRLink achieves higher recall than our proposed approach, its precision score is much lower compared to our model.
Hence, Hybrid-Linker ultimately outperforms FRLink based on F-measure which is the harmonic mean of recall and precision.
Moreover, obtaining high recall but low precision calls for manual assessment of the predictions.
That is, a developer needs to check the predicted links and remove the incorrect ones.
This adversely affects the automated feature of the approach.

Hybrid-Linker outperforms DeepLink by $50.40\%$, $26.99\%$, and $41.34\%$ regarding the average recall, precision, and F-measure. 
Previous studies have shown deep learning-based models tend to outperform classical machine learning models.
However, as shown in a study by Hellendoorn et al.~\cite{hellendoorn2017deep}, it is possible to achieve better results using simple and well-engineered approaches
compared with vanilla deep neural networks.
According to our results, we are also able to surpass DeepLink as we carefully inspect the domain of the problem, identify and incorporate more relevant information from the non-textual channel in addition to the textual information of issues and commits.
Evidently, these types of information can help boost the performance of automatic link recovery models. 
Our results are also compatible with those reported by Ruan et al.~\cite{ruan2019deeplink} where the overall recall score of FRlink is higher than DeepLink.
However, Ruan et al.~\cite{ruan2019deeplink} originally evaluated
using six projects with almost identical number of true/false links,
while in this study we have included $12$ projects with various number of true/false links and sizes to improve diversity of our dataset.
This may cause the drop in individual scores reported in this work and Ruan et al.'s~\cite{ruan2019deeplink} study (regarding comparison with FRlink).

Furthermore, our approach uses fewer computational resources and time while training the models.
For instance, pertaining the Airflow project,
the required time to train Hybrid-Linker is $25$ minutes, while it takes about $7$ hours to train DeepLink. 
able \ref{table:execution time} provides execution times per project.
DeepLink has also reported a 5X overhead comparing their approach to FRLink.
The large overhead of DeepLink is probably due to the complex nature of deep models.
Our results indicate we can train simpler models which incorporate more relevant information, thus, achieving higher accuracy and less overhead.

We also calculated the standard deviation of the F-measure for the $12$ projects in the dataset.
Taking all projects into account, the standard deviations of the F-measure are $3.01$, $3.92$, and $4.68$ for Hybrid-Linker, DeepLink, and FRLink, respectively.
That is, our approach is more stable than the other two, hence proving to be a more generalizable approach.

\begin{table}
    \caption{execution time for each project on our hardware{}}
    \centering
     \begin{tabular}{c p{8mm} c c p{8mm} c} 
     \toprule
     \textbf{Project} & \textbf{Hybrid-Linker} & \textbf{DeepLink} & \textbf{Project} & \textbf{Hybrid-Linker} & \textbf{DeepLink}\\ [0.15ex] 
     \midrule
     Beam & 35m & 19h & Flink & ~2h & ~3d\\ 
     Freemarker & 11.5s & 30m & Airflow & 25m & ~7h\\
     Arrow & 35m & ~6h & Netbeans & 7m & 25d\\
     Ignite & 22m & ~13.5d & Isis & 28m & ~23h\\
     Groovy & 54m & ~13h & Cassandra & 33m & ~6h\\ 
     Ambari & ~4h & ~7.5d & Calcite & 31m & ~6h\\ [0.15ex] 
     \bottomrule
    \end{tabular}
    \label{table:execution time}
    \vspace{-2mm}
\end{table}

\paragraph{RQ2}
\emph{How to combine the two components of the model to achieve the best outcome?}
To incorporate as much information as possible and consequently boost the performance, we propose a hybrid model of our two distinct classifiers.
To combine the predictions of the two components, we create a linear composition of their outputs. 

\begin{figure*}
    \centering
    \includegraphics[width=0.9\linewidth]{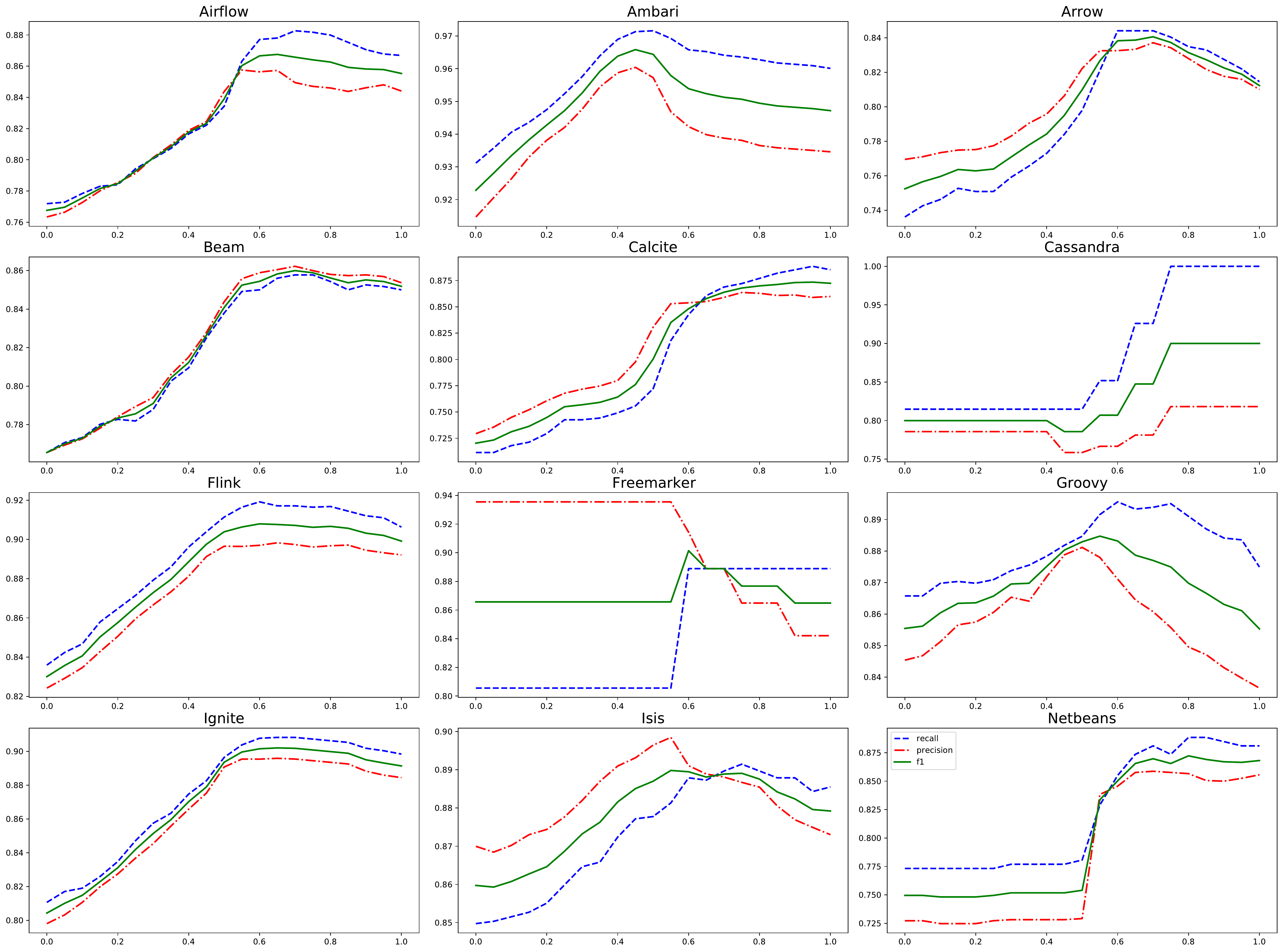}
    \caption{Tuning Alpha per project}
    \label{figure:project base alphas}
    \vspace{-5mm}
\end{figure*}

\autoref{figure:project base alphas} presents the results of using different values of alpha ranging from $0$ to $1$ for the $12$ projects under study.
As can be seen, each project requires a different value of alpha.
Thus, selecting a constant alpha for all projects will result in weaker results.

\autoref{table:best alpha project base} lists the best $\alpha$ values for each project.
In most cases, $\alpha$ is above $0.5$, with the average $\alpha$ being $0.66$ for all the projects.
This means, interestingly, in most cases the non-textual component plays a more important role in the final decision making of Hybrid-linker. 
This highlights the importance of incorporating these types of information while recovering links.
The only exception to this finding occurs in the \emph{Ambari} project with $\alpha$ of $0.45$.
This implies an approximately equal contribution of the two components of our proposed approach for this project.
On the other hand, for \emph{Calcite}, the best results are achieved with an $\alpha$ of $0.95$.
This can be a indicator that this project lacks adequate textual information useful for recovering links.
  
\begin{table} 
    \caption{Best value of Alpha per project}
    \centering
     \begin{tabular}{c c | c c | c c} 
    \toprule
     \textbf{Project} & \textbf{Alpha} & \textbf{Project} & \textbf{Alpha} & 
     \textbf{Project} & \textbf{Alpha}\\ [0.15ex] 
    \midrule
     Beam & 0.7 & Ignite & 0.65 & Flink & 0.6 \\
     Isis & 0.55 & Freemarker & 0.6 & Groovy & 0.55 \\
     Airflow & 0.65 & Cassandra & 0.75 & Arrow & 0.7 \\
     Ambari & 0.45 & Netbeans & 0.8 & Calcite & 0.95\\
    [0.25ex]
    \bottomrule
    \end{tabular}
    \label{table:best alpha project base}
\end{table}

\paragraph{RQ3}
\emph{What is the effect of each component of the model on the outcome?}
In this section, we present the results of our ablation study on assessing the effect of each component of the proposed model.
\autoref{table:textual, non-textual and hybrid comparison} presents a summary of our model's performance based on each project. 
The results indicate that on average, the performance of the textual model is lower than both the non-textual and hybrid models. 
The textual model also marks the highest standard deviation among the models with $5.83$.
Interestingly, the non-textual model outperforms the hybrid model regarding precision by $4.38\%$. 
On the other hand, the standard deviation of the non-textual model is $3.10$ which is slightly higher than the standard deviation of the hybrid model, $3.00$.
As natural language is more complex, 
text-based approaches may require more complex techniques to perform fairly good.
The higher performance of the non-textual component is probably due to 
(1) having more explicit data, 
and (2) the advantage of ensemble models.
To conclude, the hybrid model has higher recall and F-measure scores.
It also obtains the lowest standard deviation regarding its performance on all the projects.
This means, by employing both of the textual and non-textual components, the hybrid model achieves higher results, while preserving the stability of the proposed approach.
\begin{table*}
    \caption{Results of the ablation study}
    \centering
     \begin{tabular}{c | c c c | c c c | c c c} 
    \toprule
    
    \multirow{1}{*}{} &
      \multicolumn{3}{c|}{\textbf{Hybrid method}} &
      \multicolumn{3}{c|}{\textbf{Textual-based}} &
      \multicolumn{3}{c}{\textbf{Non-textual-based}}\\
    \textbf{Project} & Recall & Precision & F-measure & Recall & Precision & F-measure & Recall & Precision & F-measure \\ [0.15ex] 
    \midrule
     Beam & \textbf{85.77\%} & 86.22\% & \textbf{85.99\%} & 76.35\% & 76.35\% & 76.35\% & 84.64\% & \textbf{86.46\%} & 85.54\%\\ 
     Flink & \textbf{91.91\%} & 89.69\% & \textbf{90.79\%} & 82.52\% & 82.52\% & 82.51\% & 89.77\% & \textbf{90.32\%} & 90.05\%\\
     Freemarker & 88.89\% & \textbf{91.42\%} & \textbf{90.14\%} & 87.32\% & 88.10\% & 87.27\% & \textbf{93.93\%} & 86.11\% & 89.85\%\\
     Airflow & \textbf{87.80\%} & 85.72\% & \textbf{86.75\%} & 76.30\% & 76.30\% & 76.30\% & 81.50\% & \textbf{91.52\%} & 86.22\%\\
     Arrow & \textbf{84.41\%} & 83.71\% & \textbf{84.06\%} & 75.01\% & 75.10\% & 75.02\% & 74.98\% & \textbf{90.40\%} & 81.97\%\\
     Netbeans & \textbf{88.84\%} & 85.66\% & 87.22\% & 74.64\% & 74.76\% & 74.62\% & 82.37\% & \textbf{95.53\%} & \textbf{88.46\%}\\
     Ignite & \textbf{90.82\%} & \textbf{89.59\%} & \textbf{90.20\%} & 79.99\% & 80\% & 79.99\% & 90.07\% & 88.60\% & 89.33\%\\
     Isis & \textbf{88.13\%} & 89.84\% & \textbf{88.98\%} & 86.30\% & 86.32\% & 86.30\% & 87.08\% & \textbf{90.04\%} & 88.53\%\\ 
     Groovy & \textbf{89.15\%} & 87.79\% & \textbf{88.47\%} & 85.62\% & 85.6\% & 85.60\% & 81.20\% & \textbf{91.45\%} & 86.02\%\\ 
     Cassandra & \textbf{100\%} & 81.81\% & 90\% & 81.36\% & 81.45\% & 81.38\% & 84.37\% & \textbf{100\%} & \textbf{91.52\%}\\ 
     Ambari & \textbf{97.13\%} & \textbf{96.04\%} & \textbf{96.58\%} & 92.10\% & 92.12\% & 92.10\% & 93.37\% & 95.86\% & 94.73\%\\
     Calcite & \textbf{88.85\%} & 85.89\% & 87.34\% & 72.47\% & 72.48\% & 72.46\% & 83.55\% & \textbf{93.27\%} & \textbf{88.14\%}\\ 
    \midrule
     \textbf{Avg.} & \textbf{90.14\%} & 87.78\% & \textbf{88.88\%} & 80.83\% & 80.92\% & 80.82\% & 85.57\% & \textbf{91.63\%} & 88.36\%\\ [0.15ex] 
    \bottomrule
    \end{tabular}
    \label{table:textual, non-textual and hybrid comparison}
\end{table*}

\subsection{Discussion} \label{discussion}

Here, we present an example where our model successfully recovers the True Link between an issue and its corresponding commit.
Table~\ref{table:true link prediction example} summarizes the information of these two artifacts.
Although there are few similarities in textual information, 
the baselines and our textual component are unable to recognize this connection.
However, our non-textual component compensates for this shortcoming and predicts the correct connection. 
As it is shown, 
our model is capable of correct predictions both 
(1) when there is little textual information available
or (2) when there is no explicit relation among the text of the two artifacts.
Note that non-textual data are often available as they are automatically recorded.     

\begin{table*}
    \caption{An example of a True Link prediction}
    \centering
     \begin{tabular}{p{70mm}p{100mm}} 
     \toprule
     \textbf{Issue Information} & \textbf{Commit Information}\\ [0.25ex]
     \midrule
     \textbf{created\textunderscore date:} 2014-12-08 & \textbf{author\textunderscore time\textunderscore date:} 2014-12-10 \\ 
     \textbf{updated\textunderscore date:} 2014-12-10 & \textbf{commit\textunderscore time\textunderscore date:} 2014-12-10 \\
     \textbf{summary:} ``copy method logicalaggregate not copying indicator value properly" & \textbf{message:} ``[ calcite-511 ] copy method logicalaggregate not copying indicator value properly fixes \# 26" \\
     \textbf{description:} ``\{ \{ copy \} \} method \{ \{ logicalaggregate \} \} not take value \{ \{ indicator \} \} boolean input parameters object itself." 
     \newline
     \textbf{bug:} 1, \textbf{new feature:} 0, \textbf{task:} 0 
     & \textbf{DiffCode:} ``logical aggregate .java logical aggregate .java logical aggregate .java trait set .contains applicable convention none logical aggregate rel input immutable bit set group set immutable bit set group set aggregate call agg call logical aggregate get cluster group set logical aggregate get cluster group set group set agg call" \\
     \textbf{creator\textunderscore key:} 59f263ad6f803c44d2c8d5a716571218af230278 
     & \textbf{author:} a0465f128099ac027df4ee3910ee43aa66ad154b\\
     \textbf{closed:} 1, \textbf{open:} 0, \textbf{resolved:} 0 & \textbf{committer:} 0dc204239e76b8945e61c77525ac8f7386763a23 \\[0.25ex]
     \bottomrule
    \end{tabular}
    \label{table:true link prediction example}
    \vspace{-4mm}
\end{table*}

\subsection{Threats to Validity}\label{threat}

Here we discuss the threats to the validity of our work, 
organized into internal, external, and construct validity.

\paragraph{Internal Validity} \label{internal validity}

Internal validity is the extent to which a piece of evidence supports a claim about cause and effect, within the context of a particular study~\cite{ampatzoglou2019identifying}.
The first threat to the internal validity of our study 
is the \emph{True Link trustworthiness} and \emph{False Link trustworthiness} in our dataset.
In the case of building True Links, we have used the links provided by Claes and Mantyla in~\cite{claes202020}.
Although this dataset is validated by the authors~\cite{claes202020}, incorrect links may still be present due to human error.
Any combination other than a True Link can be considered a False Link. 
However, due to the diversity and multitude of choices for creating False Links, we had to employ several constraints as explained in Section~\ref{data selection}. 
These constraints affect our results.
According to previous studies, if an issue is related to a commit, there is a higher chance it will be answered/solved by a commit within seven days. 
Thus, by creating different combinations of False Links within seven days, we aim to create a more relevant and appropriate False Link dataset for training the models.
Lastly, data balancing is an important issue to keep in mind. 
Although one can easily create a large number of False Links, lack of enough True Links adversely affects the performance of classifiers.
To tackle this problem,
we balanced the dataset by selecting a random subset of the over-presented class before training.
Other balancing techniques are also viable.

\paragraph{External Validity} \label{external validity}

External validity is concerned with the generalizability 
of the approach and results~\cite{ampatzoglou2019identifying}.
In that regard, the dataset used in this study affects the outcome of the models.
The size and quality of the data play an important role in having a good issue and commit link predictor. 
We addressed this threat by evaluating our approach against data from multiple projects and studying the results.
As discussed in Section~\ref{results}, the lower standard deviation achieved by Hybrid-Linker indicates that results from this approach are more stable across projects.
That is, the approach is more generalizable than the state-of-the-art baselines and produces results in an expected range when applied on data from different projects.

\paragraph{Construct Validity} \label{construct validity}

Construct validity is concerned with the evaluation of the models~\cite{ampatzoglou2019identifying}.
Similar to previous work ~\cite{ruan2019deeplink,sun2017frlink,sun2017improving,schermann2015discovering,le2015rclinker}, we use precision, recall, and F-measure to evaluate the performance of our approach. 
To evaluate our proposed model more fairly, we also use five-fold cross-validation in all model evaluation steps of the study and report the average of the metrics.
By breaking data into five smaller chunks and re-evaluating the model, we ensure that all of the data has been used for training and testing.

\section{Related Work} \label{related work}

\par In this section, 
we review the related studies with the purpose of 
linking issues to their corresponding commits. 
We categorize these approaches into three major groups of 
heuristic-based, Machine-Learning-based, and Deep-learning-based studies.

\paragraph{Heuristic-based approaches}
These studies simply define 
a set of heuristics to find the links between issues and commits.
\emph{ReLink}~\cite{wu2011relink}, \emph{MLink}~\cite{nguyen2012multi}, 
and \emph{PaLiMod}~\cite{schermann2015discovering} fall into this category.
Wu et al.~\cite{wu2011relink}, introduced ReLink,
an approach that builds on top of traditional heuristics 
for creating True Links.
The traditional heuristics used in this work 
mostly rely on hints or links developers leave about bug fixes in changelogs. 
For instance, they search for keywords such as `fixed' and `bug', 
or bug ID references in changelogs.  
Moreover, they would try to find the link by using features 
extracted from linked issues and commits. 
They obtained $89\%$ precision and $78\%$ recall on average.
Nguyen et al.~\cite{nguyen2012multi} presented MLink, 
a layered approach that exploits both textual and code-related features.
They outperform ReLink by $13\%$ to $17\%$ on recall 
and $8\%$ to $17\%$ on precision~\cite{nguyen2012multi}.
However, they used only three projects when evaluating their work.
Moreover, their results showed that some individual layer's precision or recall are very low.
Finally, Schermann et al.~\cite{schermann2015discovering} introduced PaLiMod 
to enable the analysis of interlinking characteristics in commit and issue data.
They used this analysis to define their heuristics. 
PaLiMod achieves a precision of $96\%$ and recall of $92\%$ in the case of the Loner heuristic which are
single commits with no link to the addressed issues. 
Also, their method reach overall precision of $73\%$ with a recall of $53\%$ in the case of the Phantom heuristic which are
commits without a link in a series of commits that address a certain issue.
Although the idea of the Phantom case was novel, 
the results were not significant compared to former heuristic methods such as MLink.
One of the drawbacks of these studies is 
using a manually-created dataset by the authors themselves~\cite{bissyande2013empirical}.
Most of these cases used manually labeled data which reduces the confidence in the results.

\paragraph{Machine-Learning-based approaches}
 The second approach to recovering links is to use
traditional binary classifications, including \emph{RCLinker}~\cite{le2015rclinker}, \emph{FRLink}~\cite{sun2017frlink} and \emph{PULink}~\cite{sun2017improving}.
RcLinker employed \emph{ChangeScribe}, 
a tool for creating a commit message 
and used a set of features to recover the links.
They outperformed MLink in terms of F-measure by $138.66\%$~\cite{le2015rclinker}. 
ChangeScribe creates highly detailed commits 
which are not very suitable for feature extraction in this context.
Recently, FRLink was introduced which uses its own set of features~\cite{sun2017frlink}. 
The authors also use complementary documents 
such as non-source documents to learn from relevant data
They analyze and filter out irrelevant source code files 
to reduce data noise. 
FRLink outperforms RCLinker in F-measure by $40.75\%$
when achieving the highest recalls.
However, their approach encounter problems when 
(1) a dataset has a low percentage of non-source documents in commits, 
or (2) it has few or no similar code terms in the issue report and corresponding fixing commits. 
Also, text and code features were equally weighted in this approach.
A close study to FRLink is PULink~\cite{sun2017improving}, 
where authors labeled their data as a True Link/unlabeled instead of True/False Links. 
They claim they can obtain the same value of precision and recall 
with almost $70\%$ of the number of True Links in other approaches.
However, they too had a problem when a dataset has a low percentage of True Links. 
Generally, the main problem of these studies
is the low performance based on metrics like F1, precision, and recall. 
Although FRLink achieves higher recall scores,
its precision and F1 are very low.    

\paragraph{Deep-Learning-based approaches}
Xie et al.~\cite{xie2019deeplink} proposed \emph{DeepLink}~\cite{xie2019deeplink},
which incorporates a knowledge based graph and deep learning 
to solve this problem.
Using class embeddings in commit codes, the authors created this graph.
Authors also use CBOW and Word2Vec embedding for commit and issue documentation. 
As we did not have access to the knowledge graph and replication package,
we were not able to replicate this approach.
Another publication also named \emph{DeepLink}~\cite{ruan2019deeplink} 
uses a semantically-enhanced link recovery method
based on deep neural networks 
to tackle this problem. 
The authors use recurrent neural networks 
on the textual information of issues and commits
to train their model.
They disregard comments because of their length and noise. 
They have added semantic to their model to have a better prediction. 
DeepLink outperforms FRLink in terms of F-measure by $10\%$~\cite{ruan2019deeplink}. 
The challenge with deep learning algorithms 
lies in the need for a large amount of data and high computational resources.
Moreover, training these models takes a lot of time.

\par We propose a model that outperforms the baselines
by exploiting information from both textual and non-textual channels.
We use fewer resources and our training and prediction time are much lower.
We also train with projects where fewer issues and commits are available.
Thus our model will not fail when there is little historical data available for a project.

\section{Conclusion and Future Work}\label{conclusions}
The importance of recovering true connections 
between issues and their corresponding commits 
greatly affects various software maintenance tasks. 
Previous studies mostly focused on exploiting textual information 
to train their models to identify the links.
However, we introduced a hybrid method, 
called Hybrid-Linker 
based on classical ML-based classifiers,
that employs both textual and non-textual information to recover the links.
For each project, we tune alpha, 
as an indication of the importance of each information channel.
The results suggest that the non-textual information 
indeed help the predictions.
This is highlighted in cases that there is little textual information available.
Moreover, our approach requires shorter training time 
and outperforms both the competing methods, namely
DeepLink~\cite{ruan2019deeplink} and FRLink~\cite{sun2017frlink} 
by $41.3\%$ and $31.3\%$ on F-measure, respectively.

In the future, 
we plan to 
boost our proposed classifier 
by identifying new features
from different bug tracking and version control systems.
We will also investigate other classifier architectures.

\section*{Acknowledgement}
We would like to acknowledge Mahtab Nejati for her valuable comments and help on this work.

\bibliographystyle{IEEEtran}
\bibliography{main}

\begin{thebibliography}{10}
\providecommand{\url}[1]{#1}
\csname url@samestyle\endcsname
\providecommand{\newblock}{\relax}
\providecommand{\bibinfo}[2]{#2}
\providecommand{\BIBentrySTDinterwordspacing}{\spaceskip=0pt\relax}
\providecommand{\BIBentryALTinterwordstretchfactor}{4}
\providecommand{\BIBentryALTinterwordspacing}{\spaceskip=\fontdimen2\font plus
\BIBentryALTinterwordstretchfactor\fontdimen3\font minus
  \fontdimen4\font\relax}
\providecommand{\BIBforeignlanguage}[2]{{%
\expandafter\ifx\csname l@#1\endcsname\relax
\typeout{** WARNING: IEEEtran.bst: No hyphenation pattern has been}%
\typeout{** loaded for the language `#1'. Using the pattern for}%
\typeout{** the default language instead.}%
\else
\language=\csname l@#1\endcsname
\fi
#2}}
\providecommand{\BIBdecl}{\relax}
\BIBdecl

\bibitem{sun2017improving}
Y.~Sun, C.~Chen, Q.~Wang, and B.~Boehm, ``Improving missing issue-commit link
  recovery using positive and unlabeled data,'' in \emph{2017 32nd IEEE/ACM
  International Conference on Automated Software Engineering (ASE)}.\hskip 1em
  plus 0.5em minus 0.4em\relax IEEE, 2017, pp. 147--152.

\bibitem{ruan2019deeplink}
H.~Ruan, B.~Chen, X.~Peng, and W.~Zhao, ``Deeplink: Recovering issue-commit
  links based on deep learning,'' \emph{Journal of Systems and Software}, vol.
  158, p. 110406, 2019.

\bibitem{le2015rclinker}
T.-D.~B. Le, M.~Linares-V{\'a}squez, D.~Lo, and D.~Poshyvanyk, ``Rclinker:
  Automated linking of issue reports and commits leveraging rich contextual
  information,'' in \emph{2015 IEEE 23rd International Conference on Program
  Comprehension}.\hskip 1em plus 0.5em minus 0.4em\relax IEEE, 2015, pp.
  36--47.

\bibitem{anvik2006should}
J.~Anvik, L.~Hiew, and G.~C. Murphy, ``Who should fix this bug?'' in
  \emph{Proceedings of the 28th international conference on Software
  engineering}, 2006, pp. 361--370.

\bibitem{dit2013feature}
B.~Dit, M.~Revelle, M.~Gethers, and D.~Poshyvanyk, ``Feature location in source
  code: a taxonomy and survey,'' \emph{Journal of software: Evolution and
  Process}, vol.~25, no.~1, pp. 53--95, 2013.

\bibitem{sun2017frlink}
Y.~Sun, Q.~Wang, and Y.~Yang, ``Frlink: Improving the recovery of missing
  issue-commit links by revisiting file relevance,'' \emph{Information and
  Software Technology}, vol.~84, pp. 33--47, 2017.

\bibitem{claes202020}
M.~Claes and M.~V. M{\"a}ntyl{\"a}, ``20-mad: 20 years of issues and commits of
  mozilla and apache development,'' in \emph{Proceedings of the 17th
  International Conference on Mining Software Repositories}, 2020, pp.
  503--507.

\bibitem{pavlyshenko2018using}
B.~Pavlyshenko, ``Using stacking approaches for machine learning models,'' in
  \emph{2018 IEEE Second International Conference on Data Stream Mining \&
  Processing (DSMP)}.\hskip 1em plus 0.5em minus 0.4em\relax IEEE, 2018, pp.
  255--258.

\bibitem{friedman2002stochastic}
J.~H. Friedman, ``Stochastic gradient boosting,'' \emph{Computational
  statistics \& data analysis}, vol.~38, no.~4, pp. 367--378, 2002.

\bibitem{ting2011naive}
S.~Ting, W.~Ip, A.~H. Tsang \emph{et~al.}, ``Is naive bayes a good classifier
  for document classification,'' \emph{International Journal of Software
  Engineering and Its Applications}, vol.~5, no.~3, pp. 37--46, 2011.

\bibitem{breiman2001random}
L.~Breiman, ``Random forests,'' \emph{Machine learning}, vol.~45, no.~1, pp.
  5--32, 2001.

\bibitem{chen2016xgboost}
T.~Chen and C.~Guestrin, ``Xgboost: A scalable tree boosting system,'' in
  \emph{Proceedings of the 22nd acm sigkdd international conference on
  knowledge discovery and data mining}, 2016, pp. 785--794.

\bibitem{mckinney2011pandas}
W.~McKinney \emph{et~al.}, ``pandas: a foundational python library for data
  analysis and statistics,'' \emph{Python for High Performance and Scientific
  Computing}, vol.~14, no.~9, pp. 1--9, 2011.

\bibitem{aiello2016machine}
S.~Aiello, C.~Click, H.~Roark, L.~Rehak, and P.~Stetsenko, ``Machine learning
  with python and h2o,'' \emph{H2O. ai Inc}, 2016.

\bibitem{pedregosa2011scikit}
F.~Pedregosa, G.~Varoquaux, A.~Gramfort, V.~Michel, B.~Thirion, O.~Grisel,
  M.~Blondel, P.~Prettenhofer, R.~Weiss, V.~Dubourg \emph{et~al.},
  ``Scikit-learn: Machine learning in python,'' \emph{the Journal of machine
  Learning research}, vol.~12, pp. 2825--2830, 2011.

\bibitem{hellendoorn2017deep}
V.~J. Hellendoorn and P.~Devanbu, ``Are deep neural networks the best choice
  for modeling source code?'' in \emph{Proceedings of the 2017 11th Joint
  Meeting on Foundations of Software Engineering}, 2017, pp. 763--773.

\bibitem{ampatzoglou2019identifying}
A.~Ampatzoglou, S.~Bibi, P.~Avgeriou, M.~Verbeek, and A.~Chatzigeorgiou,
  ``Identifying, categorizing and mitigating threats to validity in software
  engineering secondary studies,'' \emph{Information and Software Technology},
  vol. 106, pp. 201--230, 2019.

\bibitem{schermann2015discovering}
G.~Schermann, M.~Brandtner, S.~Panichella, P.~Leitner, and H.~Gall,
  ``Discovering loners and phantoms in commit and issue data,'' in \emph{2015
  IEEE 23rd International Conference on Program Comprehension}.\hskip 1em plus
  0.5em minus 0.4em\relax IEEE, 2015, pp. 4--14.

\bibitem{wu2011relink}
R.~Wu, H.~Zhang, S.~Kim, and S.-C. Cheung, ``Relink: recovering links between
  bugs and changes,'' in \emph{Proceedings of the 19th ACM SIGSOFT symposium
  and the 13th European conference on Foundations of software engineering},
  2011, pp. 15--25.

\bibitem{nguyen2012multi}
A.~T. Nguyen, T.~T. Nguyen, H.~A. Nguyen, and T.~N. Nguyen, ``Multi-layered
  approach for recovering links between bug reports and fixes,'' in
  \emph{Proceedings of the ACM SIGSOFT 20th International Symposium on the
  Foundations of Software Engineering}, 2012, pp. 1--11.

\bibitem{bissyande2013empirical}
T.~F. Bissyande, F.~Thung, S.~Wang, D.~Lo, L.~Jiang, and L.~Reveillere,
  ``Empirical evaluation of bug linking,'' in \emph{2013 17th European
  Conference on Software Maintenance and Reengineering}.\hskip 1em plus 0.5em
  minus 0.4em\relax IEEE, 2013, pp. 89--98.

\bibitem{xie2019deeplink}
R.~Xie, L.~Chen, W.~Ye, Z.~Li, T.~Hu, D.~Du, and S.~Zhang, ``Deeplink: A code
  knowledge graph based deep learning approach for issue-commit link
  recovery,'' in \emph{2019 IEEE 26th International Conference on Software
  Analysis, Evolution and Reengineering (SANER)}.\hskip 1em plus 0.5em minus
  0.4em\relax IEEE, 2019, pp. 434--444.

\end{thebibliography}
\end{document}